\documentclass[twocolumn,showpacs,preprintnumbers, prc,superscriptaddress]{revtex4}
\usepackage{epsfig}
\usepackage{graphicx}
\usepackage{epstopdf}
\usepackage{amsmath,amssymb,amsfonts}
\usepackage{array}
\usepackage{url}
\usepackage{hyperref}
\usepackage{lineno}
\usepackage{xspace}
\usepackage[usenames,dvipsnames]{color}
\newcommand{\sqsn}{\mbox{$\sqrt{s_{_{NN}}}$}\xspace}
\newcommand{\bef}{\begin{figure}}
\newcommand{\eef}{\end{figure}}
\newcommand{\bc}{\begin{center}}
\newcommand{\ec}{\end{center}}

\newcommand{\auau}{\mbox{Au$+$Au}\xspace}

\begin{document}
\title{Higher moments of net-proton multiplicity distributions in a heavy-ion event 
pile-up scenario}

\author{P. Garg}
\email {prakhar@rcf.rhic.bnl.gov}
\affiliation{Department of Physics and Astronomy, Stony Brook University, SUNY, 
Stony Brook, New York 11794-3800, USA}
\author{D. K. Mishra}
\email {dkmishra@barc.gov.in}
\affiliation{Nuclear Physics Division, Bhabha Atomic Research Center, Mumbai 
400085, India}
\begin{abstract}
High-luminosity modern accelerators, like the Relativistic Heavy Ion Collider 
(RHIC) at BNL and Large Hadron Collider (LHC) at CERN, inherently have event 
pile-up scenarios which significantly contribute to physics events as a background. 
While state-of-the-art tracking algorithms and detector concepts take care of these 
event pile-up scenarios, several offline analytical techniques are used to remove 
such events from the physics analysis. It is still difficult to identify the 
remaining pile-up events in an event sample for physics analysis. Since the fraction 
of these events is significantly small, it may not be as serious of an issue for 
other analysis as it would be for an event-by-event analysis. Particularly, when the 
characteristics of the multiplicity distribution are observable, one needs to be very 
careful. In the present work, we demonstrate how a small fraction of residual pile-up 
events can change the moments and their ratios of an event-by-event net-proton 
multiplicity distribution, which are sensitive to the dynamical fluctuations due to 
the QCD critical point. For this study we assume that the individual event-by-event 
proton and antiproton multiplicity distributions follow Poisson, negative binomial 
or binomial distributions. We observe a significant effect in cumulants and their 
ratios of net-proton multiplicity distributions due to pile-up events, particularly 
at lower energies. It might be crucial to estimate the fraction of pile-up events in 
the data sample while interpreting the experimental observable for the critical point.

\end{abstract}

\maketitle

\section{Introduction}
\label{sec:intro} 
The recent Beam Energy Scan (BES) program performed with STAR and PHENIX detectors 
at Relativistic Heavy Ion Collider (RHIC) and the upcoming upgrades to the STAR 
experiment for BES-II are motivated to explore the phase diagram of strong 
interaction. Quantum chromodynamics (QCD) predicts a phase transition from a hadron 
gas (HG) to a quark gluon plasma (QGP) phase in the temperature ($T$) and baryon 
chemical potential ($\mu_{B}$) plane of phase diagram~\cite{Stephanov:1998dy}. 
Lattice QCD indicates a smooth crossover at ${\mu_{B} \approx 0}$, while other 
models predict a first order phase transition at higher baryon 
densities~\cite{Aoki:2006we,Alford:1997zt,Fodor:2004nz,Stephanov:2004wx,
Bazavov:2012vg}. This suggests an existence of the QCD critical end point (CEP) as 
a termination point of the first order phase transition line at finite $\mu_{B}$ and 
$T$~\cite{Ejiri:2005wq,Stephanov:1999zu}. 

The event-by-event fluctuations of conserved quantities such as, net-baryon, 
net-charge, and net-strangeness are proposed as a useful observable to find the 
existence of CEP~\cite{Koch:2005vg, Asakawa:2000wh,Asakawa:2009aj}. The correlation 
length ($\xi$) of the system is related to the moments of the multiplicity 
distribution of the above conserved quantities~\cite{Stephanov:2008qz}. Thus, these 
moments can be used to look for phase transition and the CEP by varying the 
colliding beam energy~\cite{Ejiri:2005wq,Bazavov:2012vg}. The variance $\sigma^2$ of 
these distributions is related to $\xi$ as 
$\sigma^2\sim\xi^2$~\cite{Stephanov:1999zu}. The higher order moments such as 
skewness $S$ and kurtosis $\kappa$ are even more sensitive to $\xi$ as 
$S\sim\xi^{4.5}$ and 
$\kappa\sim\xi^7$~\cite{Stephanov:2008qz,Asakawa:2009aj,Gavai:2010zn,Cheng:2008zh}. 
The higher order moments have stronger dependence on the correlation length, hence, 
these moments are even more sensitive to the dynamical 
fluctuation~\cite{Stephanov:2008qz,Asakawa:2009aj}. 
The moments (mean, $\sigma$, $S$, and $\kappa$) of the net multiplicity distribution 
are related to the cumulants ($C_n, n = 1, 2, 3, 4$)  as: mean ($M$) = $C_1$, 
$\sigma^2$ = $C_2$ = $\langle (\Delta N)^2\rangle$, $S$ = $C_3/C_2^{3/2}$ = $\langle 
(\Delta N)^3\rangle/\sigma^3$ and $\kappa$ = $C_4/C_2^2$ = $\langle (\Delta 
N)^4\rangle/\sigma^4 - 3$, where $N$ is the multiplicity of the net distribution and 
$\Delta N$ = $N-M$. The ratio of various $n$th-order cumulants $C_n$ of the 
distribution are related to the ratios and product of the moments as: $\sigma^2/M$ = 
$C_2/C_1$, $S\sigma$ = $C_3/C_2$, $\kappa\sigma^2$ = $C_4/C_2$, and $S\sigma^3/M$ = 
$C_3/C_1$. One advantage of measuring the cumulant ratios is that the volume 
dependence of individual cumulants cancels out to first order. Further, the cumulant 
ratios can be related to the ratios of the generalized susceptibilities calculated in 
lattice QCD~\cite{Ejiri:2005wq,Bazavov:2012vg,Cheng:2008zh} and other statistical 
model calculations~\cite{Karsch:2010ck}.

The measurement of net-proton~\cite{Adamczyk:2013dal,Aggarwal:2010wy} and 
net-charge~\cite{Adare:2015aqk,Adamczyk:2014fia} multiplicity distributions from BES 
at RHIC have drawn much attention from both the theoretical and experimental 
communities. There have been speculations that the non-monotonic behavior of 
$\kappa\sigma^{2}$ as a function of center-of-mass energy ($\sqsn$) in the net-proton 
multiplicity ($N_{\rm diff}$ = $N_p - N_{\bar p}$) distribution measured by the 
STAR~\cite{Adamczyk:2013dal} experiment may be an indication of the QCD critical 
point. Several studies have been carried out to estimate the excess of dynamical 
fluctuations such as the effect of kinematical acceptance~\cite{Garg:2013ata}, 
inclusion of resonance decays~\cite{Mishra:2016qyj,Begun:2006jf,Nahrgang:2014fza}, 
exact (local) charge conservation~\cite{Bzdak:2012an,Schuster:2009jv}, excluded 
volume corrections~\cite{Fu:2013gga,Bhattacharyya:2015zka} and so forth to provide a 
proper thermal baseline for experimental 
measurements~\cite{Netrakanti:2014mta,Mishra:2015ueh,Tarnowsky:2012vu,Mishra:2016ytr, 
Garg:2012nf}. 

Recently, preliminary results from the STAR experiment on the net-proton multiplicity 
distribution show a large enhancement in $\kappa\sigma^2$ values at lower collision 
energies~\cite{Luo:2015ewa}. Several theoretical studies suggest that, the higher 
moments start to oscillate with temperature and $\mu_B$ near the QCD critical 
point~\cite{Stephanov:2011pb,Schaefer:2011ex,Luo:2015doi}. The oscillating behavior 
observed in the experimental data motivated us to study the effect of residual
pile-up events. Most of the pile-up events are removed using different experimental 
techniques, however one can not rule-out the possibility of a small fraction of 
``residual" pile-up events. The residual pile-up effect has never been considered 
while studying the cumulants in the experimental data. In the present work, we 
discuss the possibility of residual pile-up events as an artifact which can be 
present in these measurements and it's influence on the results on higher moments of 
net-proton multiplicity distributions. This effect can be more pronounced at future 
heavy-ion experiments like CBM at FAIR, which will exceed collision rates up to 10 
MHz ~\cite{CBM:book}. The effect of residual pile-up events is important and should 
be considered before making any conclusion on critical point from the experimental 
data.

In high luminosity heavy-ion collisions, the contributions to background events may 
include the following~\cite{pile-up:thesis,CBM:book, 
Marshall:2014mza,Fedotov:2006hy,Drees:2001kv}:
\begin{enumerate}
     \item In-time pile-up events: If more than one collision occurs in the same 
bunch crossing in a collision of interest. This can be estimated by knowing 
the beam luminosity and collision cross-section at a particular \sqsn. For example,
the average store luminosity of $1.3\times 10^{26}$ $\mathrm{cm^{-2}s^{-1}}$ and 
the collision cross-section of 9.6 barn has been measured by the STAR experiment for 
19.6 GeV \auau running~\cite{lumin1,Brown:2011ge}. Therefore, the collision rate was 
around $1.25\times 10^{3}~\mathrm {s^{-1}}$($\approx$ 1.25 kHz). Further, the time 
difference between two bunches at RHIC is 109 ns; hence the contribution to the 
in-time pile-up events will be 109 ns $\times$ 1.25 kHz = 1.36$\times 10^{-4}$ at 
\sqsn = 19.6 GeV. Similarly, the collision rates for \auau collisions at \sqsn = 200 
GeV are around 60 kHz, which leads to $\sim$6.5 $\times 10^{-3}$ events as in-time 
pileup events.

\item Out-of-time pile-up events: If an additional collision occurs in a 
bunch crossing before and after the first collision. It may happen that the 
detectors are sensitive to several bunch crossings or their electronics integrate 
over more than the collision time period, and these collisions can affect the signal 
in a physics event. The STAR Time Projection Chamber (TPC) has a drift time of 40 
$\mu s$, which leads to additional pile-up events of about 0.05 and 2.4 events as 
out-of time pile-up for \sqsn = 19.6 and 200 GeV, respectively. It is to be noted 
that the high-resolution silicon vertex detector of STAR will further reduce the 
out-of-time pileup to almost zero.

\item Cavern background: Composed of mainly low energy neutrons and photons 
that can also cause radiation damage to detector elements and front-end electronics. 
The induced hits due to this background may increase the detector occupancy. This 
background is reduced by proper shielding of the detectors.

\item Beam  halo  events: As  heavy ions are accelerated through the collider, the 
dispersion in the beam is called the beam halo, a less dense region of ions that 
forms outside the beam and gives rise to the background events.

\item Beam gas events: Collisions that occur between the bunch and the residual 
gas inside the beam-pipe which generally occur off center in the detector. 
\end{enumerate}

In the experimental situation, several techniques are applied to reduce the 
background events. For example, while selecting good events for the physics analysis,
$z$ coordinates of the collision vertex within $\pm$50 cm for lower energies and 
$\pm$30 cm for higher collision energies are applied in STAR 
measurements~\cite{Adamczyk:2013dal}. 
This ensures the suppression of the cavern background and that events are not biased 
toward one side of the detector coverage. Similarly, out-of-time pile-up events can 
be removed by making sure that all the tracks come from the same bunch-crossing. This 
is taken care of by ensuring that the events contain data from fast signals of the 
detector (like the STAR time-of-flight detector). Further, the beam halo events are 
mostly forward focused and hence do not produce significant background. However, in 
order to remove the background events from the beam halo and the beam pipe, a cut on 
the transverse $x$--$y$ coordinate of the vertex position is applied. Further, a 
reference number for the particle multiplicity, specific to the center-of-mass energy 
is used to reject pile-up events. Also, various correlations between the global 
detector subsystems are used to remove the pile-up events. In spite of all the 
mentioned procedures, one may not assert the complete removal of pile-up events from 
the physics data sample. As an example, if two peripheral collision events happen 
within the same bunch crossing, it is difficult to identify them. This can be 
misinterpreted as a semi central collision if their vertices are not further apart 
than the vertex resolution of the detector system.

Since the fraction of these residual pile-up events would be significantly small, it 
may not be as serious of an issue for other analyses as it is the case of an 
event-by-event analysis. However, it may have serious consequences on the results of 
higher moments. For example, a small modification in the number of protons and/or 
antiprotons at the tail of the event-by-event multiplicity distribution can modify 
the results significantly, which is demonstrated in the present work. 
\begin{figure}[h]
\bc
\includegraphics[width=0.5\textwidth]{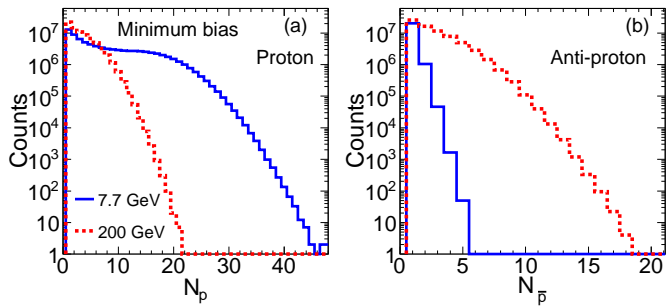}
\caption{The minimum bias proton (a) and antiproton (b) multiplicity distributions 
are shown for two different center of mass energies \sqsn = 7.7 and 200 GeV.}
\label{fig:minbias_ppbar}
\ec
\end{figure}

The paper is organized as follows: In the following section, we discuss the method 
which is used to artificially include the pile-up events. In Sec.~\ref{sec:results}, 
we show the results of net-proton multiplicity fluctuations, assuming the proton and 
antiproton multiplicity distributions as Poisson, negative binomial, and binomial 
distributions. Finally, we summarize our work and discuss its implications in 
Sec.~\ref{sec:summary}.

\begin{figure*}[ht]
\bc
\includegraphics[width=1.0\textwidth]{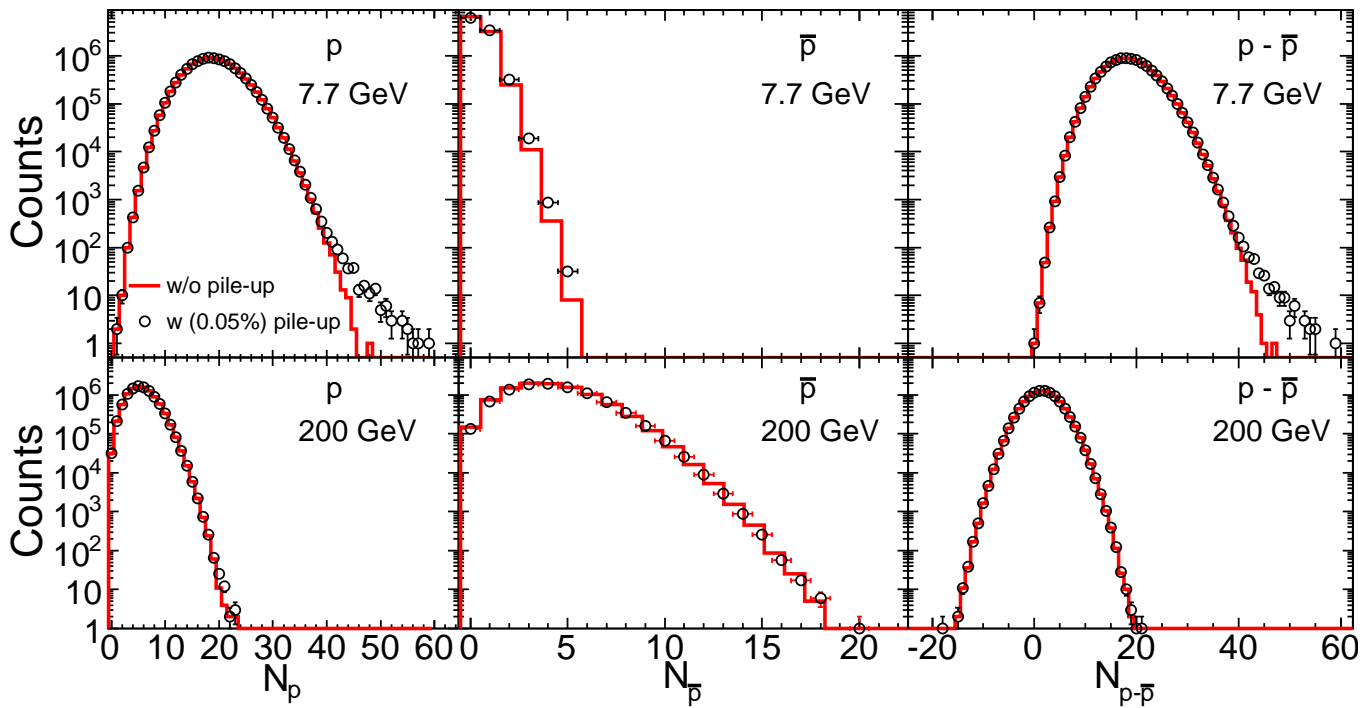}
\caption{The proton, antiproton and net-proton multiplicity distributions are shown 
with (open symbol) and without (solid line) pile-up events for $\sqrt{s_{NN}}$ = 7.7 
and 200 GeV. The proton and antiproton multiplicities from minimum bias events 
as pile-up events are added to the individual $p$ and $\bar p$ distributions, 
assuming each of the distributions are negative binomial distributions (NBD). }
\label{fig:hist_pileup}
\ec
\end{figure*}

\begin{table*}[hbt]
 \caption{Mean values of proton and antiproton distributions for most central 
(0--5$\%$) \auau collisions at various \sqsn measured by STAR 
experiment~\cite{Adamczyk:2013dal,stardata} at RHIC.}
\begin {tabular}{cccccccc}
\hline
\hline
\sqsn (GeV) & 7.7 & 11.5 & 19.6 & 27 & 39 & 62.4 & 200\\
\hline
Proton&        $18.918\pm0.009$ & $15.005\pm0.006$ & $11.375\pm0.003$ & 
$9.390\pm0.002$ & $8.221\pm0.001$ & $7.254\pm0.002$ & $5.664\pm0.001$\\
Antiproton& $0.165\pm0.001$  & $0.490\pm0.001$  & $1.150\pm0.001$  & 
$1.652\pm0.001$ 
& $2.379\pm0.001$ & $3.135\pm0.001$ & $4.116\pm0.001$\\
\hline
\hline
 \end{tabular}    
 \label{tbl:mean}
\end{table*}

\section{Method used for event pile-up studies}
\label{sec:method} 
The method discussed for this study assumes the proton and antiproton multiplicities 
to be Poisson, negative binomial or binomial distribution. As mentioned in the 
previous section, if two collision events happen within a same bunch crossing, it may 
be difficult to disentangle them and it can be misinterpreted as a single event. We 
have adopted a simple Monte Carlo approach by generating two independent multiplicity 
distributions of proton ($p$) and antiproton ($\bar p$) using the corresponding mean 
values for (0--5\%) centrality in \auau collisions at different \sqsn as given in 
Ref.~\cite{Adamczyk:2013dal}. The mean values of $p$ and $\bar p$ for different 
collision energies are also listed in Table~\ref{tbl:mean}. 

First, we assume that a large sample of central physics events has a small fraction 
of events where two central events are piled up. The extra protons and antiprotons 
coming from a certain fraction of pile-up events are added to the original 
multiplicity distribution. Hence, out of all the accumulated events, some events will 
have higher multiplicities as compared to the usual multiplicity of a central 
collision event. These high multiplicity events are distributed toward the tail of 
the distribution. The presence of a small fraction of pile-up events can have 
substantial effect on the shape of the distribution, which are described by their 
higher moments and cumulants. As a second possibility, it may also happen that an 
event from a central collision mixes with an event from another centrality class to 
form a pile-up event. In such a case, we add the $p$ and $\bar p$ multiplicities from 
a small fraction of minimum bias events to the multiplicity distribution from central 
collisions. This may be more of a probable scenario which can happen in heavy-ion 
collisions. The minimum bias distribution for protons (antiprotons) is constructed by 
combining the multiplicity of protons (antiprotons) at different collision 
centralities ranging from 0 to 80\% for each \sqsn. Figure~\ref{fig:minbias_ppbar} 
shows the minimum bias multiplicity distribution for protons and antiprotons at \sqsn 
= 7.7 and 200 GeV. Further, the multiplicity distribution of $p$ or $\bar p$ for 
different centralities are constructed using Poisson, negative binomial (NBD) or 
binomial distribution with the mean values given in Ref.~\cite{Adamczyk:2013dal}. The 
$N_{\rm diff}$ distribution is obtained on an event-by-event basis using the modified 
$p$ and $\bar p$ multiplicities.

Figure~\ref{fig:hist_pileup} shows the typical multiplicity distributions for 
protons, antiprotons, and net-protons for two different \sqsn = 7.7 and 200 GeV by 
taking their corresponding mean values. These two energies are considered to 
demonstrate the effect of pile-up event for a wider range of collision energies at 
RHIC. The multiplicity distributions are also compared with and without inclusion of 
pile-up events. In Fig.~\ref{fig:hist_pileup}, the $p$ and $\bar p$ multiplicities 
from the central events are combined with 0.05\% (five pile-up events in $10^4$ 
events) of the randomly selected $p$ and $\bar p$ multiplicities from minimum bias 
events, as shown in Fig.~\ref{fig:minbias_ppbar}. Some fraction of excess protons due 
to pile-up events can clearly be seen as compared to a purely NBD distribution at 
\sqsn = 7.7 GeV. These excess events also reflect in the $N_{\rm diff}$ 
distributions. Events with higher $p$ or $\bar p$ multiplicities will have larger 
pile-up effects, which can be observed in the proton multiplicity distribution. In 
Fig.\ref{fig:hist_pileup}, the effect of pile-up events is more visible in the proton 
distribution at \sqsn = 7.7 GeV as compared to 200 GeV. Since at lower energies the 
mean number of protons is larger as compared to higher energies, therefore, the 
effect of mixing a central event with another central (or minimum bias) event is more 
pronounced. At higher energies, due to small mean multiplicity of $p$ and $\bar p$, 
the effect does not contribute much. However, in experimentally measured $p$ and 
$\bar p$ multiplicity distributions, it is not trivial to figure out these events, as 
the excess is very small and it may look like a real event multiplicity distribution 
which can be seen for \sqsn = 200 GeV in Fig.~\ref{fig:hist_pileup}.

\section{Results and Discussions}
\label{sec:results} 
Experimentally measured $p$ and $\bar p$ distributions are usually described by 
Poisson, negative binomial or binomial distributions. Poisson expectations reflect a 
system of totally uncorrelated, and statistically random particle production. The 
Poisson statistics is a limiting case of NBD, in which both the mean and variance of 
the distribution are same. Whereas in the case of NBD, the variance is larger than 
the mean of the distribution. In case of the binomial distribution, the variance 
is less than the mean. In the present study, a range of residual pile-up event 
fraction (0.01--2\%) is considered, which may be realistic in experimental situations 
as discussed in Sec.~\ref{sec:intro}. It is to be noted that this range of pile-up 
event fractions is based on the RHIC collision rates and detector response. Further 
offline analysis techniques can further reduce this number to an even smaller 
fraction. In the following subsections, we demonstrate the pile-up effect on the 
higher moments of net-proton multiplicity distributions.

\subsection{Poisson distributions with event pile-up}
The individual proton and antiproton distributions are independently 
generated assuming each of the distribution as Poisson with the measured mean values 
as given in Table~\ref{tbl:mean}. The $N_{\mathrm{diff}}$ distribution is constructed 
by taking $N_p$ and $N_{\bar p}$ distributions on an event-by-event basis. The 
individual cumulants ($C_1$, $C_2$, $C_3$, and $C_4$) are calculated from the 
$N_{\rm diff}$ distribution for different \sqsn.

Figure~\ref{fig:cumu_poiscent} shows the collision energy dependence of cumulants 
for different fractions of pile-up events. In this case, we have added the 
multiplicities from some fraction of the central collisions as pile-up events with 
the original multiplicities from the central collisions. It is observed that, a 
small fraction of pile-up events can have a significant effect on the cumulants and 
their ratios of the net-proton multiplicity distributions. Smaller fractions of 
pile-up events have minimal effect on lower moments (cumulants) such as $M$ ($C_1$) 
and $\sigma^2$ ($C_2$) of the distribution, where as larger effects are observed for 
higher cumulants ($C_3$ and $C_4$). Figure~\ref{fig:cumu_ratios_poiscent} shows the 
ratios of cumulants as functions of \sqsn for different fractions of pile-up events. 
The $C_{32} (= C_3/C_2)$, $C_{42} (= C_4/C_2)$, and $C_{31} (=C_3/C_1)$ ratios show a 
strong dependence with energy, and the fraction of added pile-up events. Without the 
presence of pile-up events, $C_{42}$ remains constant at 1 for all \sqsn. Even a 
small fraction of pile-up events has a large effect on $C_{42}$ values. A similar 
study is performed by mixing the proton and antiproton multiplicities from minimum 
bias events as pile-up events with the $p$ and $\bar p$ multiplicities from the 
central collision events. Figure~\ref{fig:cumu_ratios_poismb} shows the collision 
energy dependence of the cumulant ratios for different fractions of pile-up events. A 
strong dependence of different fractions of pile-up events is observed particularly 
in higher cumulant ratios ($C_{32}$, $C_{42}$, and $C_{31}$). In the case of pile-up 
from minimum bias events similar qualitative behavior is observed as shown in 
Fig.~\ref{fig:cumu_ratios_poiscent}, but it is less pronounced. The effect of 
different pile-up fractions in Fig.~\ref{fig:cumu_ratios_poismb} is small as compared 
to the ones shown in Fig.~\ref{fig:cumu_ratios_poiscent} due to the type of pile-up 
events which are mixed with central events. The increase in the fraction of pile-up 
events results in higher values of cumulant ratios. As can be seen in 
Fig.~\ref{fig:cumu_ratios_poiscent}, the $C_{42}$ values increase by an order of 
magnitude at \sqsn = 7.7 GeV even in the presence of 0.1\% pile-up events.

\bef[t]
\bc
\includegraphics[width=0.5\textwidth]{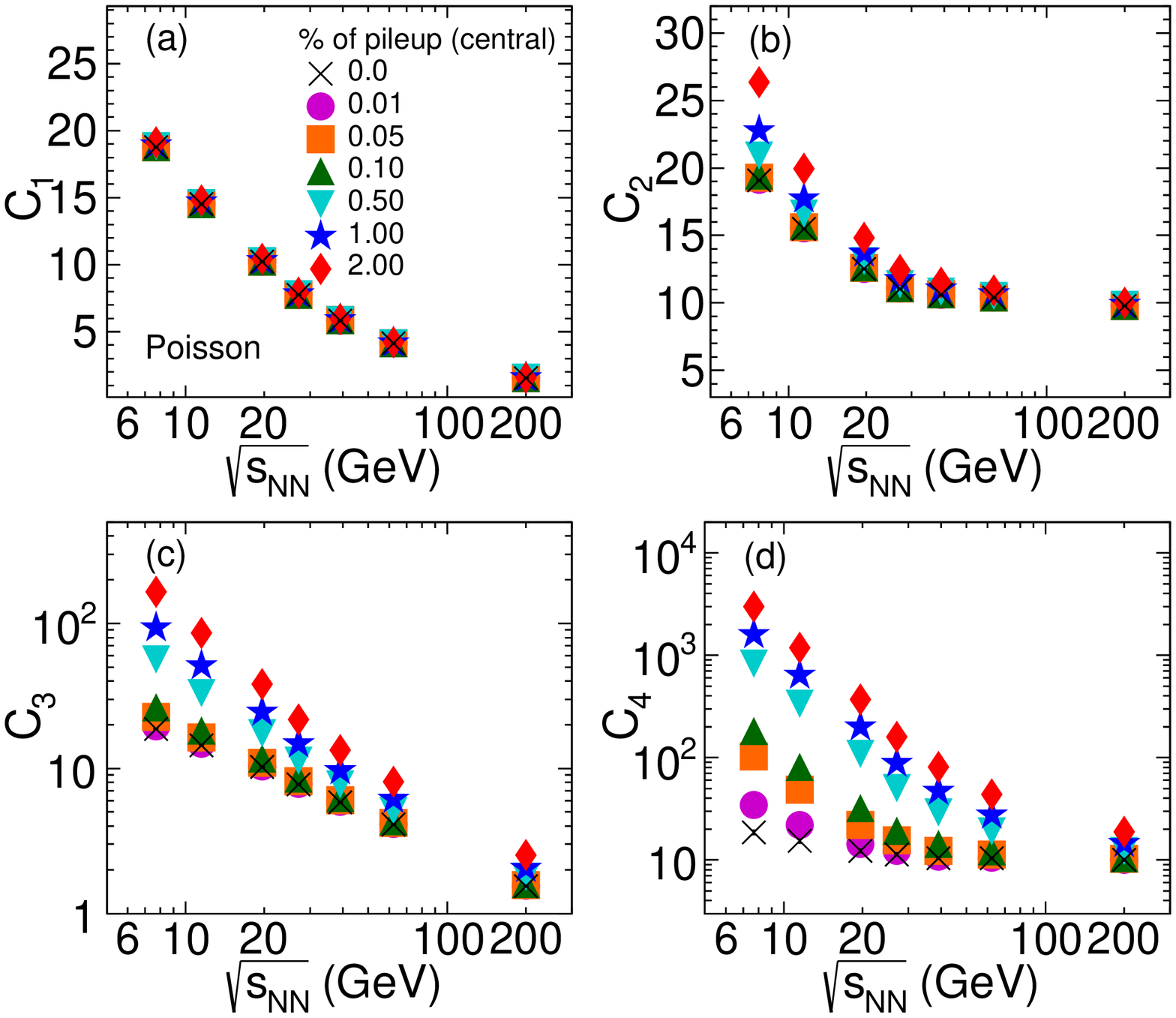}
\caption{Collision energy dependence of individual cumulants of net-proton 
distributions for different fractions of pile-up events for central (0--5$\%$) 
\auau collisions. The individual $p$ and $\bar p$ multiplicity distributions are 
assumed to be Poisson. The pile-up events from central collisions are mixed with the 
original distribution from the same centrality.}
\label{fig:cumu_poiscent} 
\ec
\eef
\bef[t]
\bc
\includegraphics[width=0.5\textwidth]{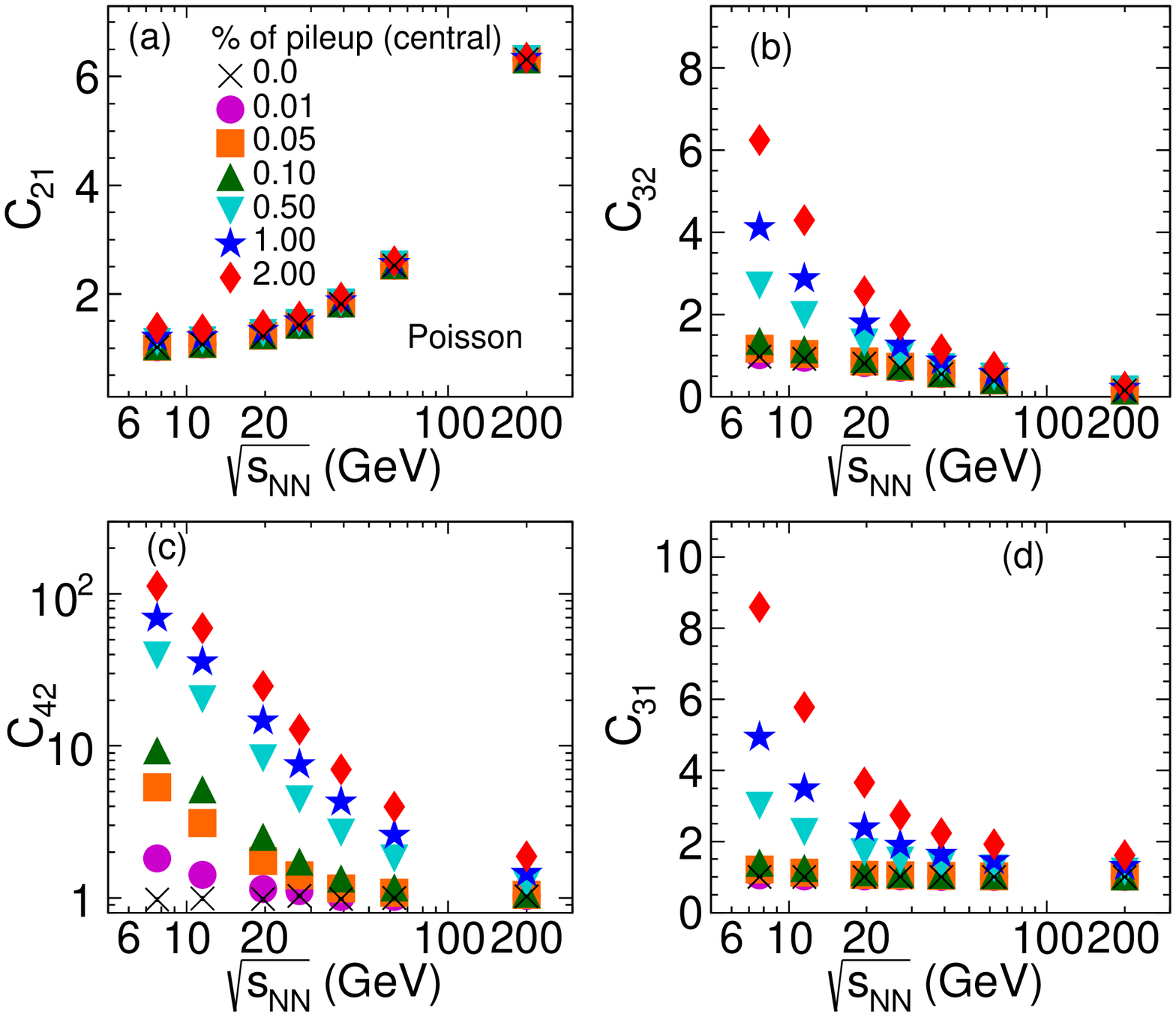}
\caption{Collision energy dependence of cumulant ratios ($C_{2}/C_{1}$, 
$C_{3}/C_{2}$, $C_{4}/C_{2}$, and $C_{3}/C_{1}$) of net-proton distributions for 
different fractions of pile-up events for central (0--5$\%$) \auau collisions. 
The individual $p$ and $\bar p$ multiplicity distributions are assumed to be 
Poisson. The pile-up events from central collisions are mixed with the original 
distribution from the same centrality.}
\label{fig:cumu_ratios_poiscent}
\ec
\eef
\bef[h!]
\bc
\includegraphics[width=0.47\textwidth]{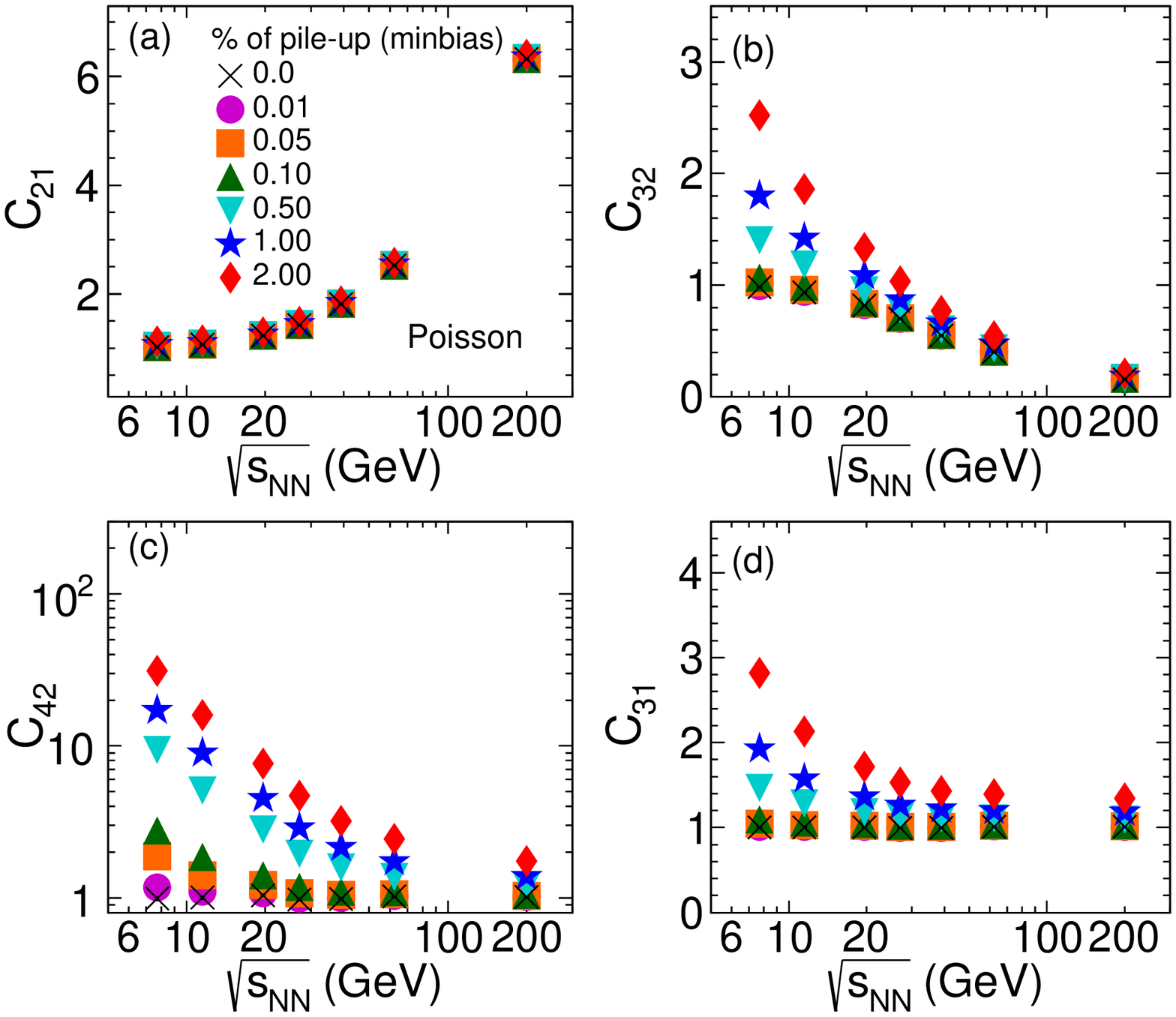}
\caption{Similar as Fig.~\ref{fig:cumu_ratios_poiscent}.
The pile-up events from minimum bias collisions are mixed with the original 
distribution from the same central collisions.}
\label{fig:cumu_ratios_poismb}
\ec
\eef

\subsection{Negative binomial distributions with event pile-up}
It is observed that particle multiplicity distributions in elementary nucleon-nucleon 
collisions, as well as heavy-ion collisions, can be well described by the negative 
binomial distribution 
(NBD)~\cite{Alner:1985zc,Abbott:1995as,Becattini:1996,Adare:2008ns}. The NBD function 
of an integer $n$ is defined as 

\begin{equation}
 NBD(n) = \frac{\Gamma(n+k)}{\Gamma(n+1)\Gamma(k)}\frac{(\langle 
n\rangle/k)^n}{(1+\langle n\rangle/k)^{n+k}}
\end{equation}
where $\langle n\rangle$ is the mean of the distribution and $k$ is an additional 
parameter. In the limiting case of $k\rightarrow\infty$, the NBD reduces to a Poisson 
distribution. The individual proton and antiproton multiplicity distributions are 
constructed assuming each distribution is a NBD with their corresponding mean values. 
The $k$ values are taken as 500 and 550 for $p$ and $\bar p$, respectively. A fixed 
$k$ value is considered for $p$ or $\bar p$ in all the \sqsn to avoid inclusion of 
additional correlation between the particles, which can change the shape of the NBD 
distribution.

The individual cumulants are calculated from the $N_{\rm diff}$ distribution, which 
is constructed by taking individual $p$ and $\bar p$ multiplicity distributions. 
Figure~\ref{fig:cumu_nbdcent} shows the \sqsn dependence of cumulants for different 
fractions of pile-up events. Both the added pile-up multiplicities and the original 
multiplicity distributions are from the central \auau collisions. Like the case of 
the Poisson distribution, the effect of the pile-up events is larger for higher 
cumulant values ($C_3$ and $C_4$). Figure~\ref{fig:cumu_ratios_nbdcent} shows the 
cumulant ratios as a function of \sqsn for different fraction of pile-up events. In 
this case also, the $C_{32}$, $C_{42}$, and $C_{31}$ ratios show strong dependence on 
energy and the fraction of added pile-up events. Figure~\ref{fig:cumu_ratios_nbdmb} 
shows the \sqsn dependence of cumulant ratios for different fractions of pile-up 
events by mixing the pile-up multiplicities from the minimum bias events with the $p$ 
and $\bar p$ multiplicity distributions from the central collisions.

\bef[t]
\bc
\includegraphics[width=0.5\textwidth]{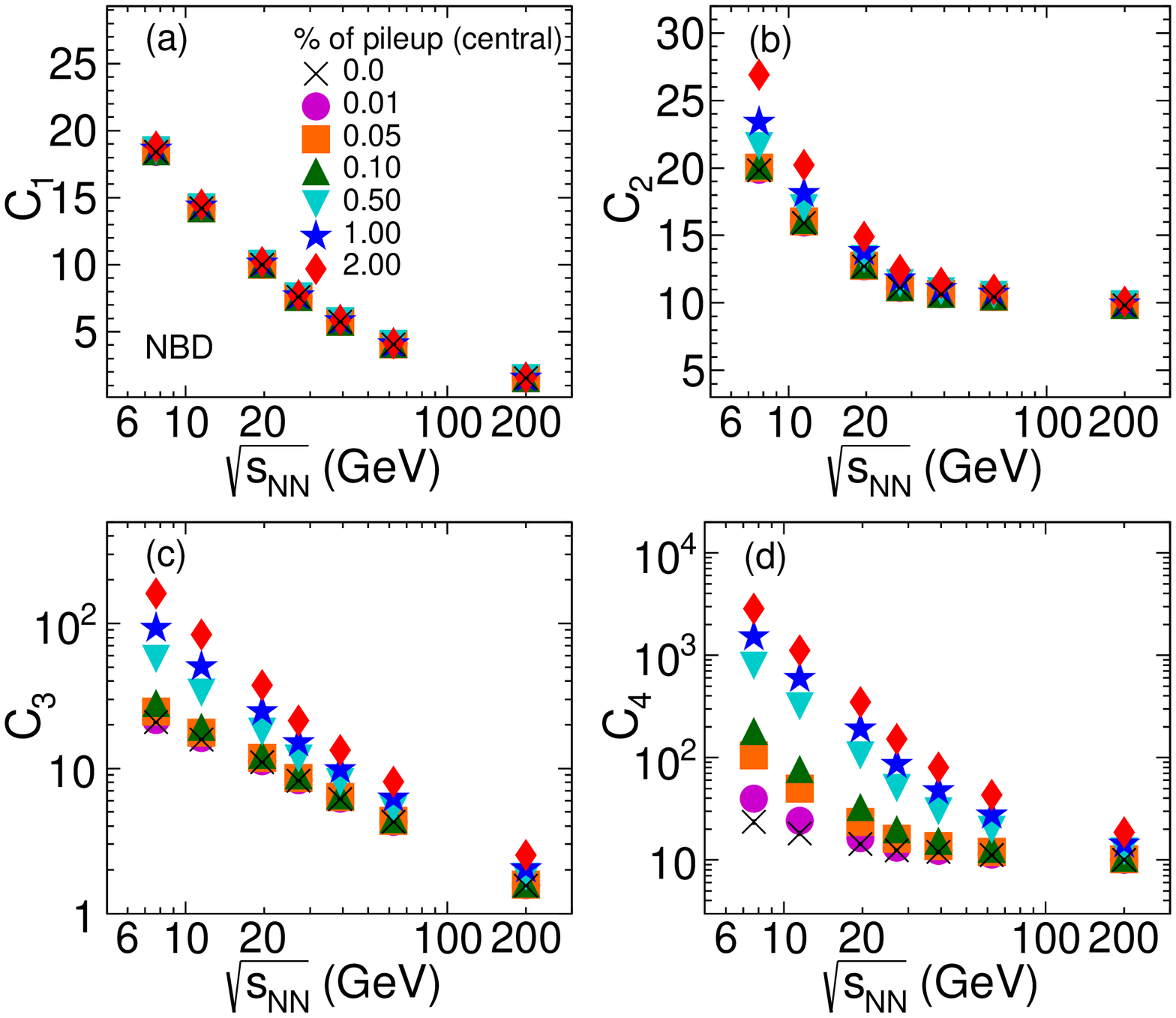}
\caption{Variation of cumulants of net-proton distributions as a function of \sqsn 
for different fraction of pile-up events for central (0--5$\%$) \auau collisions. The 
individual $p$ and $\bar p$ multiplicity distributions are assumed to be NBD. The 
pile-up events from central collisions are mixed with the original distribution from 
the same centrality.}
\label{fig:cumu_nbdcent}     
\ec
\eef
\bef[t]
\bc
\includegraphics[width=0.5\textwidth]{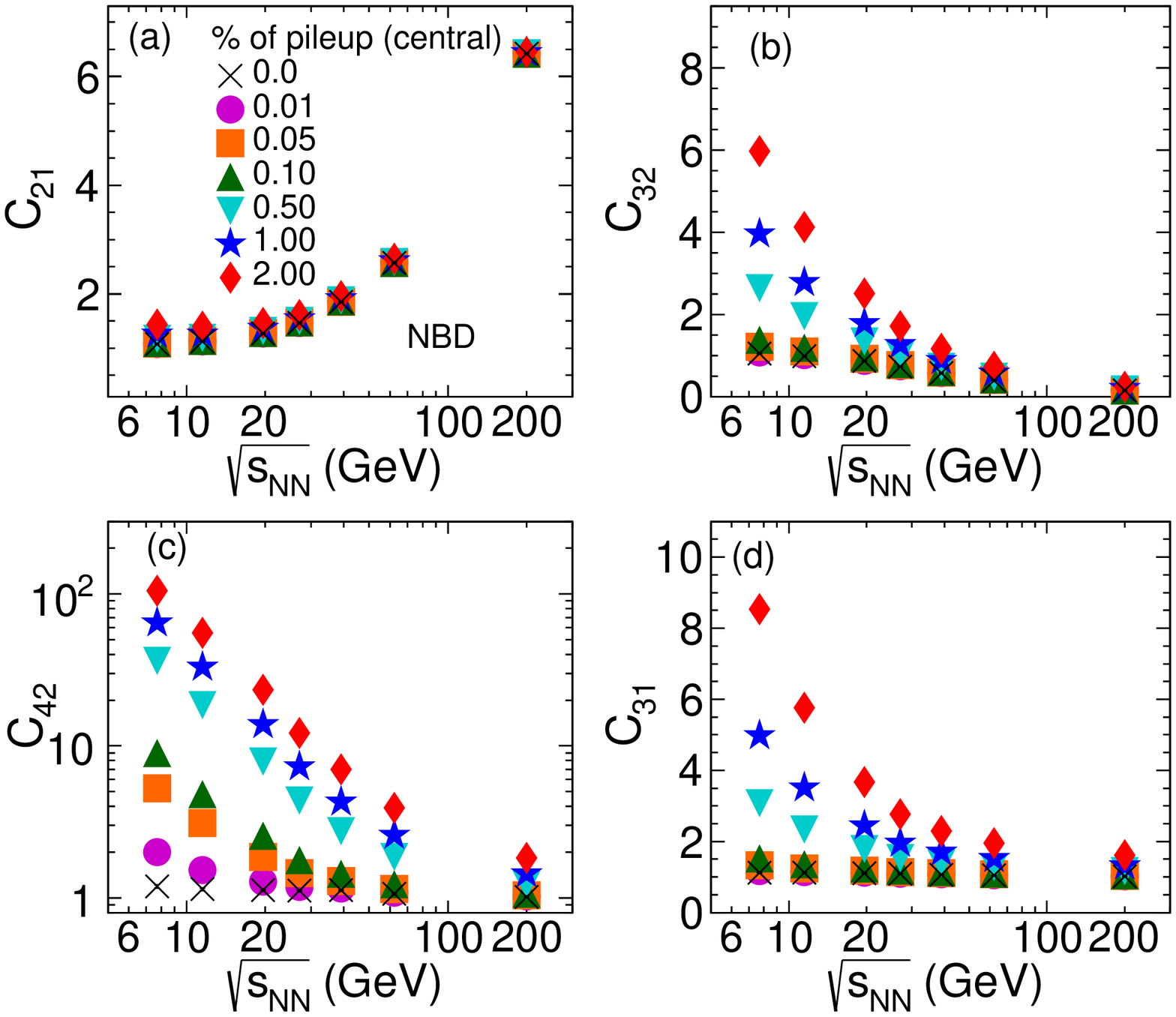}
\caption{Variation of cumulant ratios ($C_{2}/C_{1}$, $C_{3}/C_{2}$, $C_{4}/C_{2}$, 
and $C_{3}/C_{1}$) of net-proton distributions as a function of \sqsn for different 
fraction of pile-up events for central (0--5$\%$) \auau collisions. The individual 
$p$ and $\bar p$ multiplicity distributions are assumed to be NBD. The pile-up events 
from central collisions are mixed with the original distribution from the same 
centrality.}
\label{fig:cumu_ratios_nbdcent}
\ec
\eef
\bef[ht!]
\bc
\includegraphics[width=0.5\textwidth]{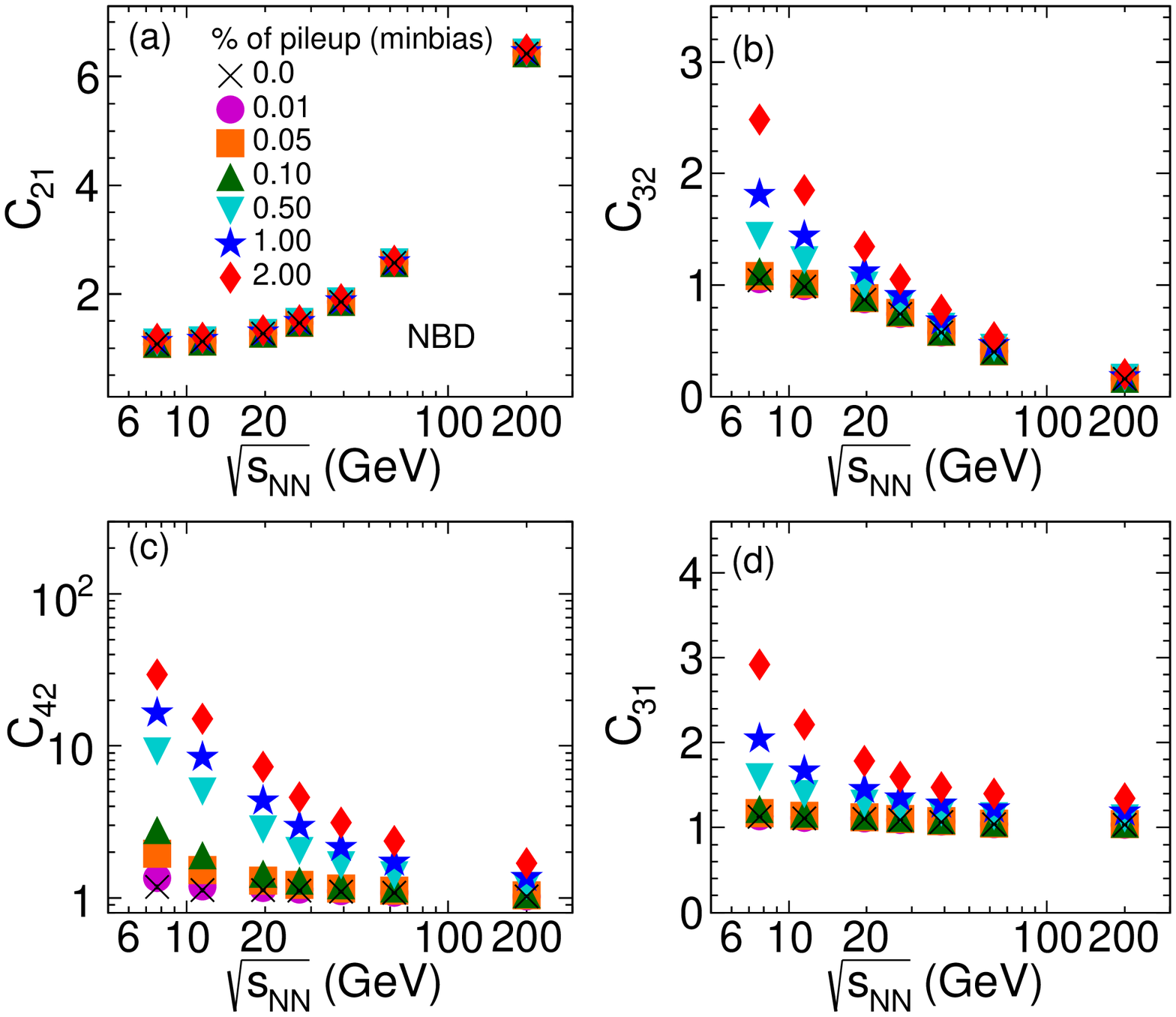}
\caption{Similar to that mentioned in Fig.~\ref{fig:cumu_ratios_nbdcent}. The pile-up 
events from minimum bias collisions are mixed with the original distribution from the 
same centrality.}
\label{fig:cumu_ratios_nbdmb}
\ec
\eef

\bef[t]
\bc
\includegraphics[width=0.5\textwidth]{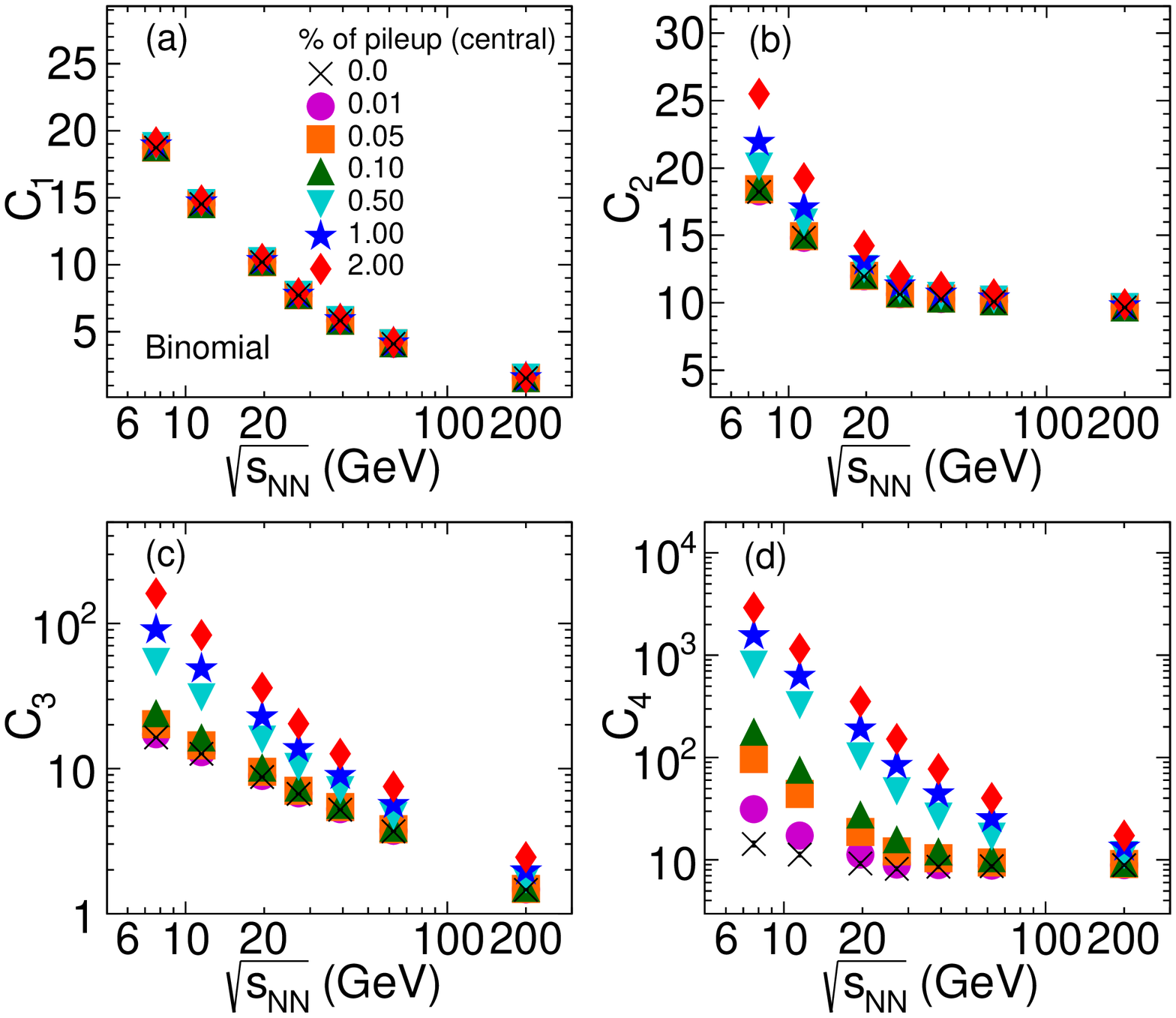}
\caption{Variation of cumulants of net-proton distributions as a function of \sqsn 
for different fraction of pile-up events for central (0--5$\%$) \auau collisions. The 
individual $p$ and $\bar p$ multiplicity distributions are assumed to be binomial. 
The pile-up events from central collisions are mixed with the original distribution 
from the same centrality.}
\label{fig:cumu_binocent}     
\ec
\eef
\bef[t]
\bc
\includegraphics[width=0.5\textwidth]{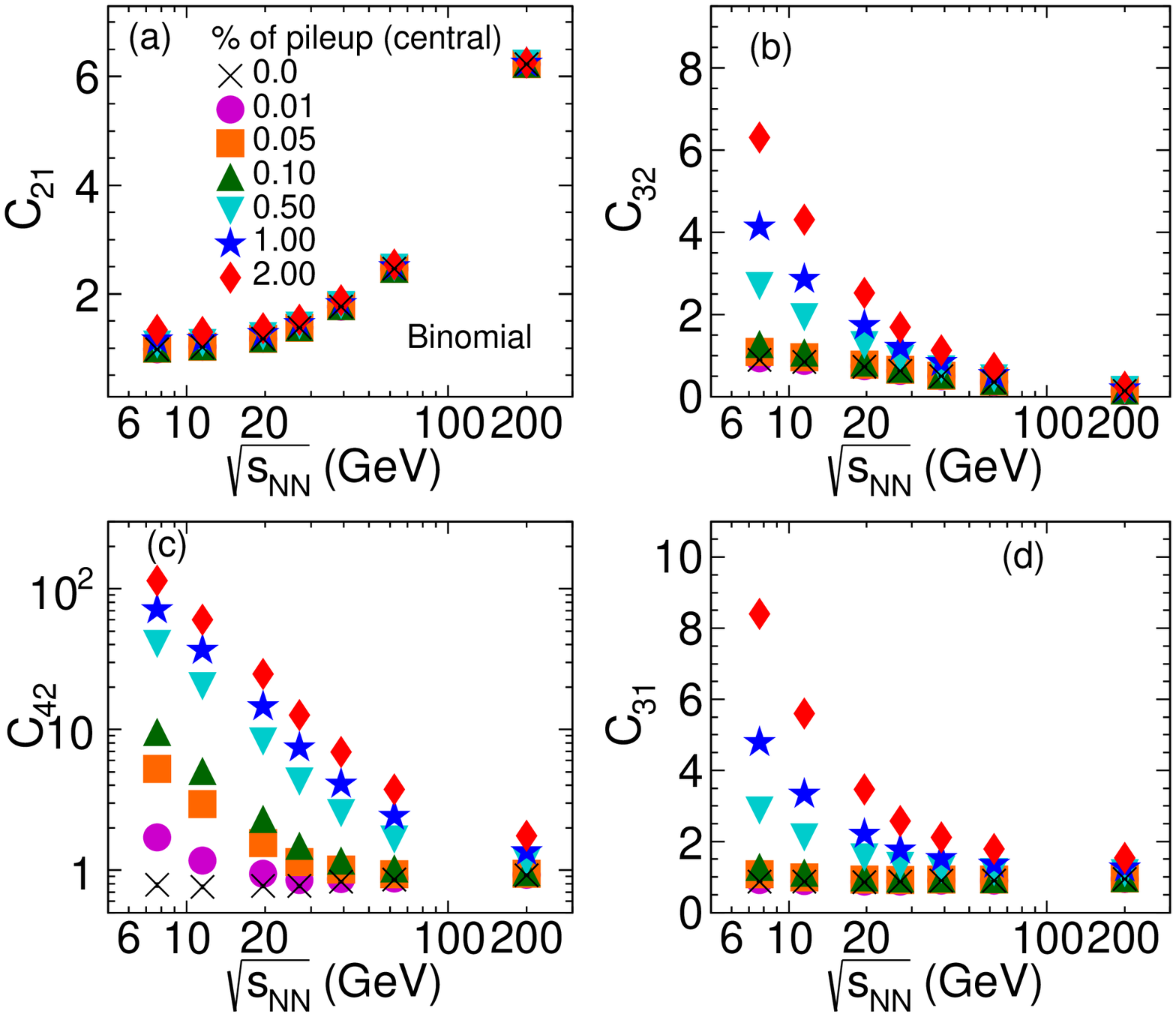}
\caption{Variation of cumulant ratios ($C_{2}/C_{1}$, $C_{3}/C_{2}$, $C_{4}/C_{2}$, 
and $C_{3}/C_{1}$) of net-proton distributions as a function of \sqsn for different 
fraction of pile-up events for central (0--5$\%$) \auau collisions. The individual 
$p$ and $\bar p$ multiplicity distributions are assumed to be binomial. The pile-up 
events from central collisions are mixed with the original distribution from the same 
centrality.}
\label{fig:cumu_ratios_binocent}
\ec
\eef
\bef[ht!]
\bc
\includegraphics[width=0.5\textwidth]{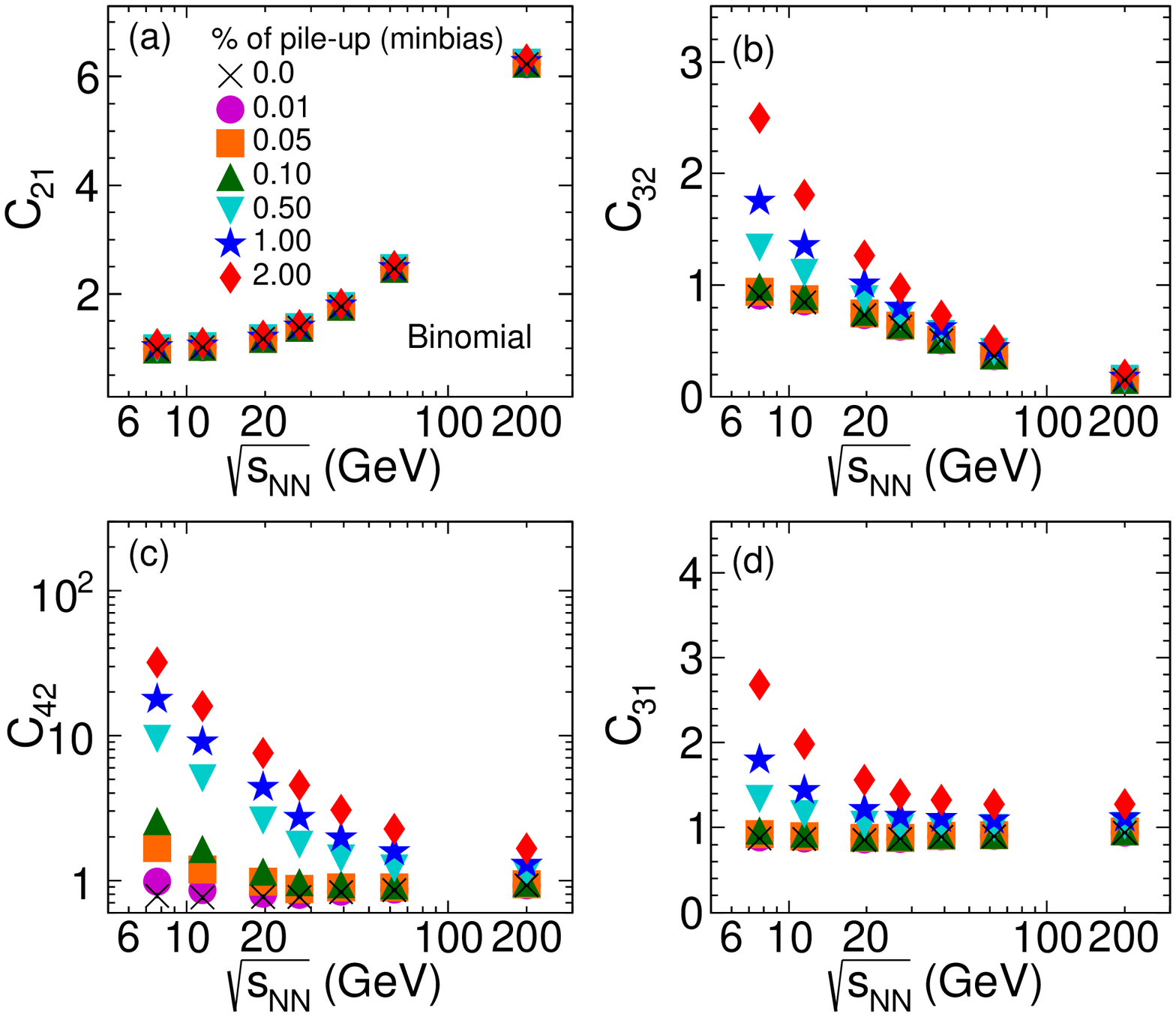}
\caption{Similar to that mentioned in Fig.~\ref{fig:cumu_ratios_binocent}. The 
pile-up events from minimum bias collisions are mixed with the original distribution 
from the same centrality.}
\label{fig:cumu_ratios_binomb}
\ec
\eef
\subsection{Binomial distributions with event pile-up}
The binomial distributions, used to explain the multiplicity distributions, are 
constructed using the mean ($C_1$) and variance ($C_2$) values of the proton and 
antiproton multiplicities as given in Refs.~\cite{Adamczyk:2013dal,stardata}. Looking 
at the individual cumulants of proton and antiproton from 
Refs.~\cite{Adamczyk:2013dal,stardata}, the individual proton and antiproton 
distributions resemble the binomial distribution. The multiplicity distributions are 
assumed to be a binomial distribution as

\begin{equation}
 B(n,p) = \frac{n!}{k!(n-k)!}p^{k}(1-p)^{n-k},
\end{equation}
where $k$ is the observed multiplicity, $n$ is the particles being produced and $p$ 
is the probability to measure it. The $C_1$ and $C_2$ are related to the above 
parameters as $C_1$ = $np$ and $C_2$ = $np(1-p)$. The net-proton distribution is 
obtained by assuming that both the proton and antiproton are produced binomially. 
Figure~\ref{fig:cumu_binocent} shows the \sqsn dependence of cumulants of net-proton 
distributions for different fraction of pile-up events. The added pile-up 
multiplicities and the original multiplicity distributions are from 0--5\% \auau 
collisions. The effects of pile-up events are larger for higher cumulants as in the 
cases of Poisson and NBD. Figures~\ref{fig:cumu_ratios_binocent} and 
~\ref{fig:cumu_ratios_binomb} show the cumulant ratios as a function of \sqsn for 
different fraction of pile-up events from central and minimum bias collisions, 
respectively mixed with the $p$ and $\bar p$ multiplicity distributions from the 
central collisions. The $C_{32}$, $C_{42}$, and $C_{31}$ ratios show strong 
dependence on energy and the fraction of added pile-up events.

In all the cases, i.e., Poisson, NBD, and binomial, the higher order cumulant ratios 
have strong dependence on the fraction of pile-up events. The effect of event pile-up 
on $N_{\rm diff}$ distribution will be more crucial at lower\sqsn, due to large 
asymmetry between proton and antiproton multiplicities. On the other hand, the event 
pile-up effect is not observed at higher collision energies, because the mean 
multiplicities of both $p$ and $\bar p$ are small and comparable. While constructing 
the net-proton multiplicity distribution, the excess pile-up effect gets neutralize 
for the high energy collisions, while at lower \sqsn this is not the case. At lower 
energies the mean multiplicity of protons is much larger than at higher energies. 
Therefore, while mixing a central event with central (or minimum bias) event, the 
effect is more pronounced as compared to higher collision energies. Recent 
preliminary results for net-proton multiplicity from the STAR 
experiment~\cite{Luo:2015ewa} observed that there is an increase in $\kappa\sigma^2$ 
($= C_{42}$) values at lower collision energies (particularly at \sqsn = 7.7 and 
11.5 GeV). The large value observed for $C_{42}$ of net-proton multiplicity 
distributions in central collisions originates partially from the efficiency 
correction. The measured uncorrected $C_{42}$ value, which would include pileup 
effects, is close to 1. Thus, any effect from pile-up events would be magnified by 
the efficiency correction. In the present analysis, we also observe an increase in 
the higher cumulant ratios due to the presence of residual pile-up events. For \sqsn 
= 7.7 and 11.5 GeV, the mean number of protons is higher as compared to other higher 
energy collisions, which causes the increase in the cumulants due to pile-up events. 
It is to be noted that, the pile-up effect will be more important for net-proton 
fluctuations as compared to net-charge fluctuations. At lower energies, the asymmetry 
between proton and antiproton multiplicities is larger, which is not the case for 
net-charge. Hence, it is important to know how much residual pile-up effect is 
present in the experimental data. One can make a more realistic estimate of the 
residual pile-up effect on the cumulant ratios by knowing the details in a real 
experimental environment. 

\section{Summary}
\label{sec:summary} 
To summarize the present work, we have emphasized the importance of residual pile-up 
events for net-proton higher moment analysis. It is demonstrated that even a small 
fraction of the pile-up events can change the higher cumulants significantly, 
especially at lower center-of-mass energies. This issue is even more important for 
the fixed target experiments like CBM where the collision rates will be even higher. 
Using a simple Monte Carlo simulation, we consider two scenarios, namely, when 
multiplicities from central collision are mixed with other central events and when 
the multiplicities from central collisions are mixed with less central events to 
mimic the pile-up scenario. In both cases, the resulting proton and antiproton 
multiplicities are modified according to the pile-up contribution and type, which are 
used to construct the event-by-event net-proton multiplicity distribution. To 
investigate the dependence on the nature of the probability distribution, the initial 
proton and antiproton distributions are assumed to be Poisson, NBD, or binomial. 
Qualitatively, all the choices show a significant increase in $C_{32}$, $C_{42}$, and 
$C_{31}$ ratios as the fraction of pile-up events is increased. The pile-up event has 
a tendency to increase the ratios of cumulants and is more significant at lower 
energies. This observation makes it critical to estimate the purity of the measured 
physics event sample for net-proton multiplicity analysis. Preliminary results from 
the STAR experiment~\cite{Luo:2015ewa} also show an increasing trend in these 
observables at lower \sqsn. The large increase in the net-proton cumulant ratios at 
lower energies from pile-up events makes it difficult to interpret the experimental 
observable for a critical point. The measurements from the STAR experiment may not 
have a significant contribution from pile-up events because of the high-resolution 
silicon vertex detector. Future high-luminosity experiments should be careful about 
the contribution of such events as it may influence the higher moment observables. It 
is important to estimate the effect of residual pile-up events before making any 
conclusion on the critical point while using higher moments of net-proton 
multiplicity distributions, as this may lead to a very different conclusion.
    
\begin{acknowledgments}
 We would like to acknowledge Volker Koch for the stimulating discussions related to
 this work. Also we would like to thank D. E. Mihalik for careful reading of the 
 manuscript.
\end{acknowledgments}

\end{document}